\documentclass[11pt]{article}
\usepackage{epsfig}
\usepackage{pifont}
\usepackage{amsbsy}

\begin{document}
\enlargethispage{3.0cm}

% \slopp

\hyphenation{Ra-cah-expr}
\hyphenation{Ra-cah-expr-}
\hyphenation{Ra-cah-expres-sion}

\title{Studies of Er ionization energy}

\author{G.\ Gaigalas, Z.\ Rudzikas and T.\ \v{Z}alandauskas\\
        \\
        Institute of Theoretical Physics and Astronomy\\
        A.\ Go\v{s}tauto 12, Vilnius 2600, Lithuania.\\
        ${}^\dagger$ e--mail: tomas@itpa.lt \\
        \\
        \\
        } 
       
\maketitle

\begin{abstract}
This work is aimed at the multiconfigurational Hartree-Fock calculations 
of the Er ionization energy. Authors have used the ATSP MCHF version in which there 
are new codes for calculation of spin-angular parts written on the basis of the 
methodology Gaigalas, Rudzikas and Froese Fischer \cite{GRFa,GRFb}, based on the second 
quantization
 in coupled tensorial form, the angular momentum theory in 3 spaces (orbital, spin and
 quasispin) and graphical technique of spin-angular integrations. They allow the study
 of configurations with open $f$-shells without any restrictions and lead to fairly 
accurate values of spectroscopic data.

\end{abstract}
\bigskip
\bigskip

\newpage

%-----------------   INTRODUCTION    --------------------------

\section{INTRODUCTION}

There is considerable interest in understanding the physics and chemistry of 
heavy atoms and ions. The main problem in investigation of such systems
is their complexity, caused by the large number of electrons and the importance of
relativistic effects. 
Therefore the detailed description of heavy atoms and ions requires the correct treatment 
of correlation effects as well as relativistic description of the system. 
The correlation effects in the theory of many-electron atoms are treated mainly by the two methods:
{\em configuration superposition} (CI) and {\em multiconfigurational Hartree-Fock} (MCHF). 
Relativistic effects are usually included adding the relativistic corrections to the 
non-relativistic Hamiltonian (for example, the Breit-Pauli approximation), or using relativistic
two component wave functions and relativistic Hamiltonian 
(for example, the Breit approximation) \cite{R97}.
The complete and rigorous treatment of both relativistic and correlation effects for heavy atoms 
and ions is, unfortunately, practically outside of today's computational possibilities.

\medskip

Nevertheless, we expect that in case of Er
%even in heavy atoms, such as lanthanides, 
relativistic effects 
as well as correlation effects between the electrons of 'inner shells' 
(core-core correlations) are the same for the neutral atom and ion. 
We assume that these effects 
(corresponding energies) cancel each other in calculation of {\em ionization energy} (IE). 
Only correlation between 'outer' (valence) electrons gives significant contribution to IE. 
And then it may be possible to get quite accurate values of ionization energies taking into
 account of them by MCHF approach. 

\medskip

This work is aimed at checking this assumption.
% for the IE of Er. 
For this purpose we perform MCHF calculations 
using the ATSP MCHF  \cite{FF2,F2000} version in which there are new codes for
calculation of spin-angular parts written on the basis of the methodology Gaigalas, Rudzikas 
and Froese Fischer \cite{GRFa,GRFb}, based on the second quantization in coupled tonsorial form, 
the angular momentum theory in 3 spaces (orbital, spin and quasispin) and graphical 
technique of spin-angular integrations. They allow the study of configurations with 
open $f$-shells without any restrictions.

%\newpage

% -------------------   APPROACH  -------------------------

\section{APPROACH}

Ionization energy we define as $IE = E_{ion} - E_{g}$, where 
$E_{g}$ and $E_{ion}$ are the ground state energies of neutral 
and singly ionized atom correspondingly. The ground state of 
neutral Er is 
%\medskip
\begin{equation}
1s^{2}2s^{2}2p^{6}3s^{2}3p^{6}3d^{10}4s^{2}4p^{6}4d^{10}5s^{2}5p
^{6}4f^{12}6s^{2}~~{}^{3}H~\equiv [Xe]4f^{12}6s^{2}~~ {}^{3}H
\end{equation}

\medskip

and that of singly ionized Er  
%\medskip
\begin{equation}
1s^{2}2s^{2}2p^{6}3s^{2}3p^{6}3d^{10}4s^{2}4p^{6}4d^{10}5s^{2}5p
^{6}4f^{12}6s^{1}~~{}^{4}H\equiv [Xe]4f^{12}6s^{1} ~~ {}^{4}H.
\end{equation}

%--- Hamiltonian ---
\subsection{Hamiltonian}

We calculate the ground state energies making use of the Hamiltonian

\begin{equation}
{\cal H}={\cal H}_{NonRel}+{\cal H}_{RelCor},
\end{equation}

where ${\cal H}_{NonRel}$ is the usual non-relativistic Hamiltonian
and  ${\cal H}_{RelCor}$ stands for relativistic corrections.
In our calculations we separate relativistic corrections into following three 
parts:

\begin{equation}
\label{RelCor}
{\cal H}_{RelCor} = {\cal H}_{Sh}+{\cal H}_{mp}+{\cal H}_{OO}
\end{equation}

with
% mp
{\em mass-polarization} correction given by the Hamiltonian
\begin{equation}
\label{mp}
{\cal H}_{mp} = -\frac{1}{M} \sum_{i < j =1}^{N}\left({\bf p}_i\cdot 
{\bf p}_j\right),
\end{equation}

% OO
{\em orbit--orbit} term given by
\begin{equation}
\label{eq:oo}
{\cal H}_{OO}=-\frac{\alpha ^2}2\sum_{i<j=1}^N\left[ 
\frac{\left({\bf p}_i\cdot 
{\bf p}_j\right) }{r_{ij}}+
\frac{\left({\bf r}_{ij}\left( {\bf r}_{ij}\cdot {\bf 
p}_i\right)
{\bf p}_j\right) }{r_{ij}^3}\right] ,
\end{equation}

and the remaining part of relativistic corrections 
% Sh
\begin{equation}
\label{Sh}
{\cal H}_{Sh} = {\cal H}_{MC} + {\cal H}_{D1}+ {\cal H}_{D2} 
+{\cal H}_{SSC},
\end{equation}
consisting of 
  % MC
the {\em mass correction} term
\begin{equation}
{\cal H}_{MC}=-\frac{\alpha ^2}8\sum_{i=1}^N{\bf p}_i^4 ,
\end{equation}
  % D1, D2 
as well as the contact interactions, described by the one-- and two--body 
{\em Darwin terms} ${\cal H}_{D1}$ and ${\cal H}_{D2}$. They are

\begin{equation}
{\cal H}_{D1}=\frac{Z \alpha ^2 
\pi}2\sum_{i=1}^N{\boldmath{\delta }}\left(
\vec{r}_{i}\right) \qquad \mbox{and}\qquad {\cal H}_{D2} = 
-\pi \alpha ^2\sum_{i<j=1}^N{\boldmath{\delta }}\left( 
\vec{r}_{ij}\right) .
\end{equation}
  % SSC
The last addant in Eq. (\ref{Sh}) stands for the {\em spin--spin contact} term
\begin{equation}
{\cal H}_{SSC}=-\frac{8\pi \alpha ^2}3\sum_{i<j=1}^N\left( {\bf 
s}_i\cdot 
{\bf s}_j\right) \delta \left( {\bf r}_{ij}\right).
\end{equation}

The corrections presented in Eq.(\ref{RelCor}) are (except {\em mass polarization} (\ref{mp})) 
of the order of 
square of fine structure constant. They enable us to make a study of 
contribution of the main relativistic corrections to the calculations of ionization energy.  
Let us also mention that the Hamiltonian is presented in atomic units.

%--- mchf ---
\subsection{MCHF}

For calculation of ionization energy we used MCHF method. 
In this approach, the wave function is expressed as a linear 
combination of {\em configuration state functions} (CSFs) 
which are antisymmetrized products of one-electron spin-orbitals. 
A set of orbitals, or {\em active set}
(AS), determines the set of all possible CSFs or the {\em complete 
active space} (CAS) for MCHF calculation. The size of the latter grows rapidly 
with the number of electrons and also with the size of the orbital AS. 
Most MCHF expansions are therefore limited to a {\em restricted active 
space} (RAS) \cite{FF2}. 
No 'relaxation' effects were included.

% ---------   BASES - 			RAS    construction   ----------

\section{RAS construction}

Large scale systematic MCHF calculations of IE of lanthanides has not been done yet. 
Therefore, following the methodology of \cite{FF2}, it is important to investigate 
the structure of Er and Er$^{+}$ ground configurations, to impose the {\em core} and 
{\em valence} shells and to 
evaluate {\em valence--valence} (VV), {\em core--valence} (CV) and  {\em core--core} (CC) 
correlations.

It is always a question when we can assume that a subshell is a part of the core, and
when it should be treated as a valence shell. The answer is not trivial even for
boron like ions, and in our case it is even more complicated because of complexity of
configurations of Er and Er$^{+}$, and our attempt is to take care of the correlation effects
that do not cancel each other between ion and atom.

Because we treat IE in non-relativistic approach, and in the neighbourhood of the ground
level there are no levels with the same L, S, J values, methodics 
based on the consideration of energy spectra described in \cite{FF2} 
could not be straightforward adapted to impose core and valence shells in our case.  
     
Therefore in this chapter we will study some possibilities of RAS construction.

\subsection{HF calculations}

First insight into the structure of Er and Er$^{+}$ ground states we can get from 
the {\em Hartree-Fock} (HF) calculations. The resultant ground state energies and 
mean distances of $nl$ radial functions are presented in TABLE I.

\begin{table}\centering{ \textbf{
{\large TABLE I. Results of HF calculations. Values of mean distance 
from the nucleus \textless r\textgreater ~ and energies of 
ground states (in a. u.)}}}\\
\vspace{5 mm}
%\begin{center}
\begin{tabular}{ c c c} \hline \hline
 Function & \textless r\textgreater$_{Er}$ & \textless r\textgreater$_{Er^{+}}$ \\
\hline \hline
\\[-0.2cm]
$1s$ & 0.022 & 0.022 \\
$2s$ & 0.094 & 0.094 \\
$2p$ & 0.080 & 0.080 \\
$3s$ & 0.242 & 0.242 \\
$3p$ & 0.232 & 0.232 \\
$3d$ & 0.205 & 0.205 \\
$4s$ & 0.545 & 0.545 \\
$4p$ & 0.557 & 0.557 \\
$4d$ & 0.588 & 0.588 \\
$5s$ & 1.371 & 1.385 \\
$5p$ & 1.563 & 1.565 \\
$4f$ & 0.754 & 0.754 \\
$6s$ & 4.630 & 4.093 \\
\hline
%\hline
\\[-0.2cm]
Energy: & -12498.1528 & -12497.9809 \\
[0.2cm]
\hline
\hline
\end{tabular}
\end{table}

Resultant energies are in agreement with those presented in \cite{TSSMK}. The important 
note is that $6s$ function is much more remote frome the nucleus than the ones of 
$5s$, $5p$ and $4f$. And the open $4f$ shell is closer to the nucleus than the $5s$ and $5p$.    
Therefore, we have a difficulty in treatment of 'outer' electrons: usually as 
outer (valence) shells the open ones are considered, but sometimes the closed shells 
are included too \cite{FF2}. 
For light atoms these shells are spartially 'outer'.

\subsection{CORE I} 

In this case we use the core 

$I$ = $[Xe]$ ${}^{1}S$ 

and the $4f$, $6s$ we treat as valence shells. 
The $4f$ shell we treat as valence shell because it is open and $6s$ because 
the corresponding radial function in much more remote from the nucleus than others. 
This approach is close to the advices given in \cite{FF2}. 

The basis for the MCHF expansion was formed using the CSF's of configurations made 
of single and double (S, D) excitations from the valence shells to some {\em destination set}.
There were two types of destination sets used:

%\medskip

\begin{equation}
\label{core_a}
a =  \left \{ 5d,5f,5g,6p,6d\right\},
\end{equation}

%\medskip

\begin{equation}
\label{core_b}
b = a + \left \{6f,6g,6h,7s,7p,7d\right\}.
\end{equation}

%\medskip

Further on we denote the basis as core with subscript of destination set.
For example, $I_{a}$ denotes the basis, consisting of CSF's of configurations, 
made by S, D excitations from  $4f^{12}6s^{2}$ for Er and $4f^{12}6s^{1}$ for $Er^{+}$ to 
the destination set '$a$' and cores $[Xe]$.  
The numbers of CSFs in the bases are presented in TABLE II.  

The weight for the main CSF was found to be 0.977 for $I_{a}$ (and similar for $I_{b}$). 
This value is close to that (0.949) found by CI method \cite{SST}.  
The mean distances of radial functions from the nucleus are found to be smaller than for HF
calculations. 
For example \textless r\textgreater$_{4f}$ = 0.752 a.u. for $I_{a}$  (0.748 a.u. for $I_{b}$)
and \textless r\textgreater$_{6s}$ = 4.550 a.u. for $I_{a}$  (4.534 a.u. for $I_{b}$).

\subsection{CORES II, III} 

\medskip

In this case, we treat as valence shell only $6s$, because of its spatial location. 
We expect this strategy to be more efficient for the calculations of $6s$ ionization energy
because as we can see from HF calculations mean distance of $4f$ radial functions is not 
much different for Er and Er$^{+}$. As a cores we use

% cia dar darasyt apie tai kodel mes nuspr. naudot sita core.

$II$. $[Xe]4f^{12}$ with not fixed term 

and

$III$. $[Xe]4f^{12}$ with fixed term $^{3}H$. 

There were five types of destination sets used with these cores: 

(\ref{core_a}) and (\ref{core_b}) as for core I and three more

\begin{equation}
\label{core_c}
c = b + \left \{7f,7g,7h,7i,8s,8p,8d\right\},
\end{equation}

\begin{equation}
\label{core_d}
d = c + \left \{ 8f,8g,8h,8i,8k,9s,9p,9d\right\}, 
\end{equation}

\begin{equation}
\label{core_e}
e = d + \left \{9f,9g,9h,9i,9k,9l,10s,10p,10d\right\}.
\end{equation}

As we can see from TABLE II, the basis formed with the same destination sets is 
the biggest for the core I, the medium for core II and the smallest for core III.

The weights of main CSFs in MCHF expansions are about 0.960 -- 0.980 for all bases with 
cores II, III.
The mean distance from the nucleus for $6s$ radial function is greater than the one
obtained from HF calculations but smaller than obtained using bases with core I.
For example,  \textless r\textgreater$_{6s}$ = 4.560 a.u. for $III_{a}$, 4.564 a.u. 
for $III_{b,d,e}$.

% -------------    RESULTS  ------------------

%\section{RESULTS}

\begin{table}\centering{ \textbf{
{\large TABLE II. Results of MCHF calculations. Numbers of CSFs (NCSF) and values of IE (in eV.)}
%\\
%\vspace{3 mm} 
%{\footnotesize See page \ldots for Exmplanation of Tables.}
}}\\
\vspace{5 mm}
%\begin{center}
\begin{tabular}{ l l l l l } \hline \hline
 $Basis$ & NCSF (Er) & NCSF (Er$^{+}$) & $IE_{NonRel}$ & $IE_{Rel}$ \\
\hline \hline
\\[-0.2cm]
$I_{a}$ &  2838 &  2769 & 5.563 & 5.739 \\
[0.1cm]
$I_{b}$ & 12811 & 12054 & 5.572 & 5.807 \\
[0.3cm]
$II_{a}$ &   236 & 8 & 5.895 & 6.640 \\
[0.1cm]
$II_{c}$ &  2600 & 23 & 5.793 & 5.877 \\
[0.1cm]
$II_{d}$ &  5565 & 32 & 5.793 & 5.874\\
[0.1cm]
$II_{e}$ & 10347 & 43 & 5.792 & 5.877 \\
[0.3cm]
$III_{a}$ &   70 &  4 & 5.896 & 6.073\\
[0.1cm]
$III_{b}$ &  272 &  7 & 5.796 & 5.856\\
[0.1cm]
$III_{c}$ &  733 & 11 & 5.792 & 5.876\\
[0.1cm]
$III_{d}$ & 1569 & 15 & 5.792 & 5.873\\
[0.1cm]
$III_{e}$ & 2938 & 20 & 5.792 & 5.877\\
[0.2cm]
\hline
%\hline
\\[-0.2cm]
\multicolumn{3}{l}{Non relativistic HF} & 4.677 & \cite{SST}  \\
[0.1cm]
\multicolumn{3}{l}{CI} & 5.077 & \cite{SST}   \\
[0.1cm]
\multicolumn{3}{l}{CI$_{Q}$ with Davidson Q correction \cite{Dav}} & 5.250 & \cite{SST}  \\
[0.1cm]
\multicolumn{2}{l}{Estimated} & 5.58 & & \cite{SST}  \\
[0.2cm]
\multicolumn{2}{l}{Experiment} & 6.108 & & \cite{MZH}  \\
[0.2cm]
\hline
\hline
\end{tabular}
\end{table}

% -------------    CONSIDERATION   --------------------------------

\section{$6s$ IONIZATION ENERGY}

The results of MCHF calculations are presented in TABLE II. 

For the bases obtained using $II$, $III$ cores in non-relativistic approach, we get
that increasing destination set the value of IE decreases until certain value
(in our case 5.792 eV). This value should be considered 
as the result of 'frozen core' method.
The result that for certain core using smaller (for example 'a') destination set 
and correspondingly smaller basis we obtain the IE value closer to experimental one 
is treated by us as casual, because in smaller destination set (basis) we account for
smaller part of valence correlations or (and) take into account it with different precision
for Er and Er$^{+}$. 

\medskip 

Adding the relativistic corrections raise the value of IE up to 5.877 eV. This value 
is fairly close to the experimental one 6.108 eV (the error is less than 
4\%). 

\bigskip

Using core $I$ the corresponding basis is 10 times larger than the one of $II$ 
(or about 40 times larger than formed using core III). Nevertheless, the 
results obtained using these bases are much worse. For example, 
in non-relativistic approach IE value for the $I_{b}$ basis is 5.572 eV, whereas the 
corresponding value for the $I_{b}$ is 5.796 eV.  
It is due to the fact that basis formed using destination set 'b' for the 
core $I$ is not enough to account for the correlation effects of $4f$--electrons, which,
represented in full, cancels between Er and Er$^{+}$.
Relativistic corrections improve the value of IE, but it is still worse than
those obtained using cores $II$, $III$. 
Of course, the results obtained using core I could be improved using larger bases
(and should overtake the ones obtained using $II$, $III$ cores), but at present 
it is unreachable because of our computational resources. 

So, we recomend to use core III for the calculations of IE. In this case it is possible
to get quite good value of IE taking into account VV (6s--6s) correlations
only. 
The inclusion of CV$_{4f}$ and CC$_{4f}$ correlations requires much more 
computational resources and with the present ones doesn't lead to the better result.
  
%  -----   CI       ----------------------

\bigskip

And finally, let us compare our results with the previous calculations
of IE, where correlation effects were treated using CI method \cite{SST}. 

The authors used well-temped Gaussian type functions (GTF's) and augmented diffuse $p$--, 
$d$-- and $f$--functions to describe the $6s$--$6s$, $6s$--$4f$ and $6s$--$5d$ electron 
correlations as well as $8g$ and $7h$ to describe the angular correlation effects. 
The total number of GTF's was ($29s,25p,22d,17f,8g,7h$).
Since authors could not perform full single- and single-double excitation configuration 
interaction (SDCI), important CSF's were selected by performing the second order 
perturbation calculations, reducing number of CSF's. They performed so-called natural 
orbital iterations to obtain compact CI expansions for the ground state as well 
as for the ionized state. The error of the correlation energies due to unselected CSF's was 
estimated to be 0.001 a.u. and the error in the IE was estimated to be less than 0.05 eV.   
%The weight of the reference function was 94.9 \%.
As one can see from the TABLE II, the result of CI calculation was 5.077 eV, and 
the one with Davidson correction \cite{Dav} CI$_{Q}$ 5.250 eV. 
The result with estimated relativistic effects is 5.58 eV.

Comparing the results \cite{SST} with ours, (5.792 eV for non-relativistic value of IE), 
we can see that it is possible to obtain much better value of IE only accounting for
the valence ($6s$--$6s$) correlation, whereas incomplete inclusion of correlation effects 
of inner shells just disimproves results.

% -------------    CONCLUSION   --------------------------------

\section{CONCLUSION}

For the calculations of Er ionization energy the most efficient 
strategy is to use the MCHF expansions with frozen core $[Xe]4f^{12}$ $^{3}H$ and 
single, double excitations from (6s). CV$_{4f}$ and CC$_{4f}$ correlations are not important
in this case.

\medskip

Our results on erbium are more accurate than data found using the Davidson $CI_{+Q}$
method \cite{Dav} and adding the relativistic corrections \cite{SST}.  

\medskip

The results obtained show that if the correlation effects of inner shells cancel each 
other between atom and ion, then it is possible to get quite accurate data by MCHF method 
accounting for correlation effects of outer electrons only. And this assumption takes 
place in the case of Er ionization energy.  

\medskip

The results demonstrate the ability of the approach by Gaigalas et al.\cite{GRFa,GRFb} 
to obtain fairly accurate data for heavy atoms and ions, having open $f$-shells.

%\newpage

\end{document}